\documentclass[journal]{IEEEtran}
\pdfoutput = 1

\usepackage[utf8]{inputenc}
\usepackage{amsmath}
\usepackage{fixltx2e}
\usepackage{graphicx, caption}
\usepackage{color}
\usepackage[margin=20mm]{geometry}
\usepackage{amssymb}
\usepackage{hyperref}
\usepackage{amsfonts}
\usepackage{subfigure}
\usepackage{titlesec}
\usepackage{float} 
\usepackage{indentfirst}
\usepackage{mathtools}
\usepackage{array}
\usepackage{upgreek}

\usepackage{tikz}
\usetikzlibrary{calc}
\usepackage[american]{circuitikz}
\usepackage{pgfplots}
\pgfplotsset{compat=1.17}
\usetikzlibrary{external}
\tikzset{external/mode=graphics if exists}

\begin{document}
%
\title{Double Upconversion for Superconducting Qubit Control realised using Microstrip Filters}
%
%
%

\author{Jonathan Dearlove,
        Prasanna Pakkiam,
        and Arkady Fedorov \\ \emph{ARC Centre of Excellence for Engineered Quantum Systems, School of Mathematics and Physics, The University of Queensland, Saint Lucia, Queensland 4072, Australia}}
\maketitle

\begin{abstract}
Superconducting qubits provide a promising platform for physically realising quantum computers at scale. Such devices require precision control at microwave frequencies. Common practice is to synthesise such control signals using IQ modulation, requiring calibration of a in-phase (I) and quadrature (Q) signals alongside two DC offsets to generate pure tones. This paper presents an economic physical implementation of an alternative method referred to as double upconversion which requires considerably less hardware calibration and physical resources to operate a qubit. A physical circuit was created using standard PCB design techniques for microstrip filters and two common RF mixers. This circuit was then utilised to successfully control a superconducting transmon qubit. When using proper RF shielding, qubit tones were demonstrated with over 70\,dB of spurious-free dynamic range across the entire operational spectrum of a transmon qubit.
\end{abstract}


%
\IEEEpeerreviewmaketitle

\section{Introduction}
%
%
%
%

\IEEEPARstart{S}{uperconducting} qubits provide a promising hardware platform for realisation of quantum computers. Such devices are operated at microwave frequencies with arbitrary waveforms. Common practice to synthesise these control waveforms is through the use of IQ modulation, upconverting a control pulse from its synthesised frequency to the qubit frequency. This paper introduces an alternative upconversion scheme realised using two mixing stages and bandpass microstrip filters.

\begin{figure}[htb]
\centering
\begin{tikzpicture}
	\draw (0.5,0) node[left]{$\omega_{\textnormal{IF}}$} -- (1,0) node[mixer,anchor=west](mixA){};
	\draw (mixA.east) to[bandpass,>,l={$\omega_1=\omega_\textnormal{LO1}\pm\omega_\textnormal{IF}$}] (4,0)
		-- (4,0) node[mixer,anchor=west](mixB){};
	\draw (mixB.east) to[bandpass,>,l={$\omega_2=\omega_{\textnormal{LO2}}\pm\omega_1$}] (7,0) node[](out){};
	\node [anchor=west,align=center] at (out) {$\omega_2$};
	
	\draw (1.5,-1) node[below]{$\omega_{\textnormal{LO1}}$} -- (mixA.south);
	\draw (4.5,-1) node[below]{$\omega_{\textnormal{LO2}}$} -- (mixB.south);
	
	\node [inputarrow,anchor=tip] at (mixA.west) {};
	\node [inputarrow,rotate=90,anchor=tip] at (mixA.south) {};
	\node [inputarrow,anchor=tip] at (mixB.west) {};
	\node [inputarrow,rotate=90,anchor=tip] at (mixB.south) {};
	\node [inputarrow] at (out) {};
\end{tikzpicture}
\captionsetup{width=.8\linewidth}
\caption[IQ Mixing Topology]{\textbf{Circuit topology of an IQ mixer}. The ordered pairs denote the respective frequency and phase. The scheme takes in an RF tone at $\omega_\textnormal{LO}$ and splits it to form an in-phase and $90^\circ$ quadrature phase components. These components are mixed with IF signals provided by 2 AWG outputs. By carefully selecting $\phi_Q=\phi\pm\pi/2$, it is easy to show that the output tone will be at $\omega_\textnormal{LO}\pm\omega_\textnormal{IF}$. Therefore, this scheme requires two arbitrary wave generator inputs (AWGs), one local oscillator (LO) and a specialised IQ mixer device (that is, the $90^\circ$ phase-shifter, the two mixers and the adder).}
\label{figure:IQ_mixer_archi}
\end{figure}
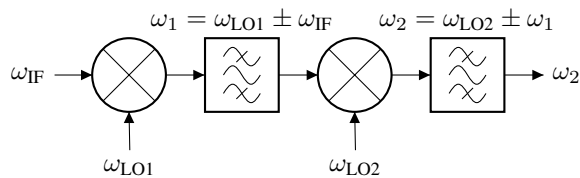

Control of a superconducting qubit is commonly performed with two separate arbitrary wave generators (AWGs) to provide an I and Q pair along with an RF source providing the local oscillator (LO) and IQ mixer. IQ modulation creates relatively pure tones (required for high-fidelity qubit operations) at the desired frequency by performing suppression of the upmixing sidebands and LO using a combination of DC offsets and phase calibration \cite{Baur2012RealizingQG}. Such a circuit that provides this functionality is presented in Figure~\ref{figure:IQ_mixer_archi}. Insufficient  suppression of unwanted side-bands in the mixing stages will result in low fidelity control as the qubit dynamics are highly sensitive to unwanted local tones and noise in the spectrum.

Utilisation of IQ modulation for qubit control introduces a variety of parameters: two AWG channels, two DC Channels and one LO channel. The benefits of IQ modulation upconversion is the range of frequencies at which signals can be synthesised. However the modulation scheme requires a more expensive IQ mixer and also comes with the burden of requiring calibration of these parameters for each frequency of operation. The focus of this paper is an alternative upconversion scheme, known as double upconversion, that simplifies calibration and reduces the number of variable parameters. In addition, double upconversion trades one expensive AWG channel for a cheaper LO channel. That is, it only requires a single AWG source and two LO channels to upconvert a control signal to the desired qubit frequency. This paper will cover the design of such a circuit followed by the successfully implementation of double upconversion to control a superconducting transmon qubit.

\section{Design}

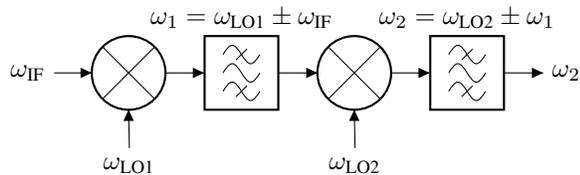
\begin{figure}[tb]
\centering
\begin{tikzpicture}
	\draw (0.5,0) node[left]{$\omega_{\textnormal{IF}}$} -- (1,0) node[mixer,anchor=west](mixA){};
	\draw (mixA.east) to[bandpass,>,l={$\omega_1=\omega_\textnormal{LO1}\pm\omega_\textnormal{IF}$}] (4,0)
		-- (4,0) node[mixer,anchor=west](mixB){};
	\draw (mixB.east) to[bandpass,>,l={$\omega_2=\omega_{\textnormal{LO2}}\pm\omega_1$}] (7,0) node[](out){};
	\node [anchor=west,align=center] at (out) {$\omega_2$};
	
	\draw (1.5,-1) node[below]{$\omega_{\textnormal{LO1}}$} -- (mixA.south);
	\draw (4.5,-1) node[below]{$\omega_{\textnormal{LO2}}$} -- (mixB.south);
	
	\node [inputarrow,anchor=tip] at (mixA.west) {};
	\node [inputarrow,rotate=90,anchor=tip] at (mixA.south) {};
	\node [inputarrow,anchor=tip] at (mixB.west) {};
	\node [inputarrow,rotate=90,anchor=tip] at (mixB.south) {};
	\node [inputarrow] at (out) {};
\end{tikzpicture}
\captionsetup{width=.8\linewidth}
\caption[Double Up Conversion Topology]{\textbf{Design topology of double up conversion scheme}. The input IF frequency $\omega_\textnormal{IF}$ provided by a single AWG output is first upconverted by $\omega_\textnormal{LO1}$ to create two tones at $\omega_\textnormal{LO1}\pm\omega_\textnormal{IF}$. A bandpass filter centred at $\omega_1$ will select one of these sidebands. The second upconversion stage at a higher $\omega_\textnormal{LO2}$ creates two tones at $\omega_\textnormal{LO2}\pm\omega_1$, whereupon one selects one of the sidebands by centering the bandpass filter at $\omega_2$ to only let through one of these tones. Thus, this design utilises two mixers with two local oscillators (LO) to convert a single drive signal generated by an AWG to the desired qubit frequency $\omega_2$.}
\label{figure:double_up_topology}
\end{figure}

Double upconversion consists of two RF mixers and two analog filters (in this case both being bandpass) as presented in Figure~\ref{figure:double_up_topology}. Such a circuit only requires a single AWG channel and two local oscillators to convert from the base band AWG frequency to the desired qubit frequency.

The operation of the double upconversion design consists of the following steps:

\begin{enumerate}
  \item The first RF source LO1 will be set at some frequency $\omega_\textnormal{LO1}$ between the AWG frequency and the qubit frequency. This creates two sidebands at frequencies $\omega_\textnormal{LO1} \pm \omega_\textnormal{IF}$.
  \item The $\omega_\textnormal{LO1}$ is carefully tuned once upon construction of the circuit such that only one of the sidebands from the first mixing stage enters the passband of the first stage bandpass filter.
  \item The single tone from the first stage bandpass filter is then upconverted via a second RF source set at frequency $\omega_\textnormal{LO2}$. The frequency $\omega_\textnormal{LO2}$ can be chosen such that only the upper or lower sideband enters the passband of the second stage bandpass filter.
  \item Thus, the final output frequency is: $\omega_\textnormal{LO2} \pm (\omega_\textnormal{LO1} \pm \omega_\textnormal{IF})$. Thus, while LO1 is set once upon construction, the frequency of LO2 can be adjusted accordingly to synthesise the drive at the desired frequency when operating a given transmon qubit. Unlike IQ modulation double up conversion also requires no re-calibration for any change in drive frequency, it only requires the second LO to be shifted.
\end{enumerate}
Standard benchtop AWGs are often limited to sampling rates between 1 and 2\,GS/s, resulting in most synthesised signals being in the range of 300-750\,MHz. Thus, if there were only a single stage upconversion, in order to have a fixed filter to attenuate the sidebands, the dynamic range of possible output frequencies would be limited to 600-1500\,MHz. To broaden this dynamic range, the first stage upconversion moves the frequency tone to several gigahertz before upconverting again to the final desired frequency.

In order to successfully attenuate unwanted tones after the first mixer stage (that is, the first stage LO frequency and the unwanted sideband), the first stage filter must have a relatively small passband in the order of several 100\,MHz to isolate the desired sideband. 
The spacing of the sidebands after the second stage mixer will be substantially large as a result of the first stage mix up being in the order of 2-3\,GHz. This much larger spacing allows for the second stage filter to have a much wider passband. This then grants the freedom to place the second stage LO above or below the passband of the second stage filter and is the primary reason for incorporating the first stage into the design. The second filter's passband can then in theory be as wide as $2 \times (\omega_\textnormal{LO1} + \omega_\textnormal{AWG})$ allowing for a large range of qubit frequencies to be synthesised.

\section{Implementation}

In order to physically realise the circuit presented in Figure~\ref{figure:double_up_topology}, custom RF band pass filters would need to be developed. The circuit in question would need to be able to convert AWG signals in the frequency range of 300-500\,MHz to qubit frequencies between 4.5-6\,GHz. In this case micro strip filters were produced for both the first and second stage filters. The chosen dielectric of FR4 (relative permittivity between 4.2 and 4.5) was chosen due to its standardisation, availability and cost. Designs were completed around the selection of this dielectric. Each PCB consists of two copper layers, one being the ground reference plane and the other being the microstrip geometry that forms the filter. All PCBs were manufactured to the standard 1.6mm thickness.  To avoid any potentially unwanted filter responses, no solder mask was used.

\subsection{Filter Design Parameters}

\begin{figure}[htb]
\centering
\includegraphics[scale = 0.24]{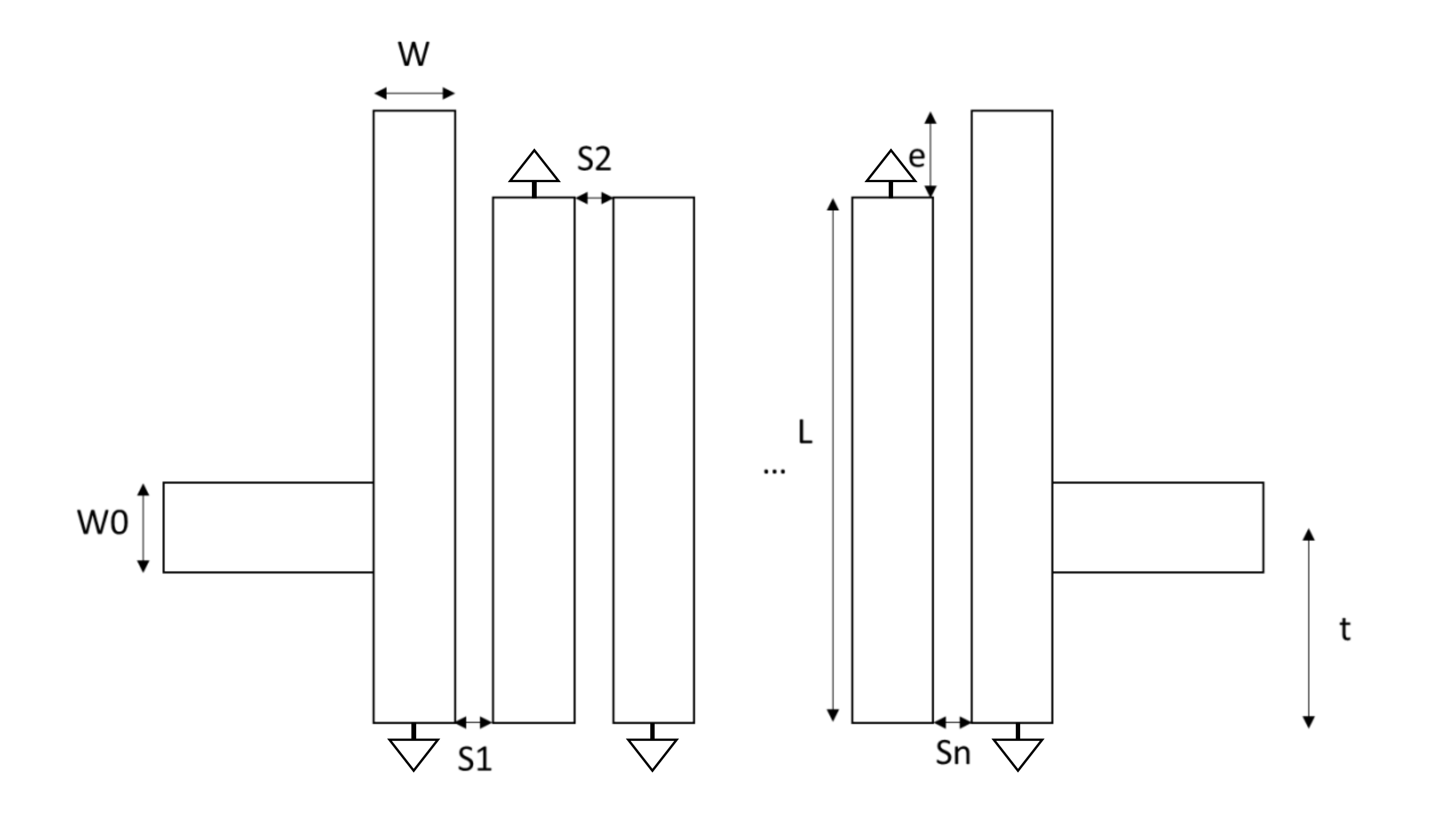}
\captionsetup{width=.75\linewidth}
\caption[Parameterisation of Interdigital Geometry]{\textbf{Interdigital filter geometry}. Filter design parameterised by length of each resonator $\{ l_1, l_2, ..., l_n\}$ and corresponding widths $\{ W_1 W_2, ..., W_n\}$. Coupling between resonators is done by fringing fields across the gaps separated by ${s_{i,i+1}}$ for $i=1, ...,n-1$. Resonators are grounded (vias to ground plane below PCB) on alternating ends. Characteristic admittance $Y_t$ of the filter can be set by the geometry of the input lines. It is common for the filter's geometry to be symmetrical along the y-axis centred halfway through the middle resonator.}
\label{figure:interdigital_param}
\end{figure}

\begin{figure}[htb]
\centering
\includegraphics[scale = 0.24]{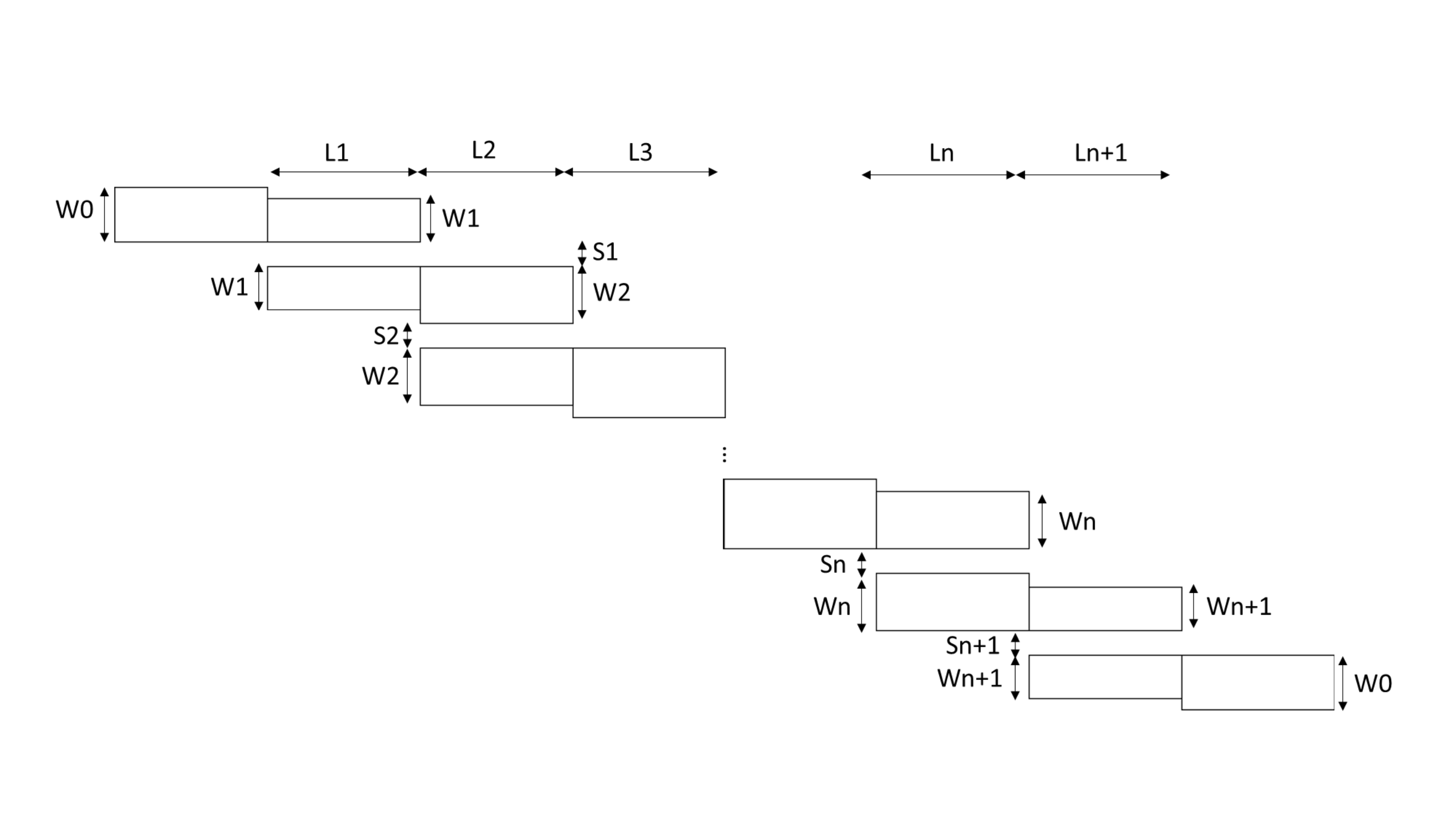}
\captionsetup{width=.75\linewidth}
\caption[Parameterisation of Parallel Coupled Geometry]{\textbf{Parallel coupled filter geometry}. Filter design parameterised by length of each half-wavelength resonator $\{ l_1, l_2, ..., l_n\}$ and corresponding widths $\{ W_1 W_2, ..., W_n\}$. Coupling between resonators is done by fringing fields across the gaps defined by distances ${s_{i,i+1}}$ where $i=1, ...,n-1$. There is a ground plane below the PCB. The parallel arrangements provide large coupling for given spacing's, resulting in larger bandwidths. Image taken from \cite{microstrip_textbook}.}
\label{figure:parallel_coupled_param}
\end{figure}

An interdigital geometry was utilised for the first stage filter for its ability to create thinner pass-bands with better roll-off at lower frequencies (in the range of 500\,MHz-4\,GHz). A Parallel geometry was selected for the second stage filter for its ability to produce a much wider passband around the desired qubit frequencies~\cite{indira_khan_nalini_2013,microstrip_textbook}. The topologies and their corresponding design parameters used in fabrication are presented in Figures \ref{figure:interdigital_param} and \ref{figure:parallel_coupled_param}, and tables \ref{table:ID12_params} and \ref{table:ID09_params}. Note that the design parameters were chosen to not only achieve the required frequency passbands but also maintain ease of mass manufacturing with the smallest feature size being 0.2\,mm.

\begin{table}[htb]
\centering
\begin{tabular}{|l|l|}
\hline
\textbf{Parameter} & \textbf{Value {[}mm{]}} \\ \hline
W0        & 1.9            \\ \hline
W         & 2.995          \\ \hline
L         & 11.07          \\ \hline
t         & 3.516          \\ \hline
e         & 0.7986         \\ \hline
S1        & 1.752          \\ \hline
S2        & 3.019          \\ \hline
S3        & 3.019          \\ \hline
d         & 1.6          \\ \hline
\end{tabular}
\captionsetup{width=.8\linewidth}
\caption[Interdigital Filter parameters]{\textbf{Interdigital filter design parameters}. Parameters match those shown in figure~\ref{figure:interdigital_param} with d being the PCB thickness. This fifth order filter was designed for a passband of 500\,MHz between 3 and 3.5\,GHz.}
\label{table:ID12_params}
\end{table}

\begin{table}[htb]
\centering
\begin{tabular}{|l|l|}
\hline
\textbf{Parameter} & \textbf{Value {[}mm{]}} \\ \hline
W1        & 1.321          \\ \hline
W2        & 1.359          \\ \hline
W3        & 1.667          \\ \hline
I1        & 6.702          \\ \hline
I2        & 6.691          \\ \hline
I3        & 6.611          \\ \hline
S1        & 0.2            \\ \hline
S2        & 0.2            \\ \hline
S3        & 0.2            \\ \hline
d         & 1.6          \\ \hline
\end{tabular}
\captionsetup{width=.8\linewidth}
\caption[Parallel Filter parameters]{\textbf{Parallel filter  design parameters}. Parameters match those shown in Figure~\ref{figure:parallel_coupled_param} with d being the PCB thickness. This fifth order filter was designed for a passband between 4.5 and 8\,GHz.}
\label{table:ID09_params}
\end{table}

\subsection{Simulation of Filter Designs}

In order to verify the design processes of microstrip filters used in the upconversion process, computational simulations were developed and run using COMSOL. These simulations allowed for analysis of the electromagnetic modes and scattering parameters present across the microstrip filters. These simulations are performed in 3D and were setup within a bounded area with a height of roughly 10 times the substrate height above the surface of the filter. This bounding represents the final box that the filters will be placed in. In these simulations the ports were set to $50\,\Omega$ impedance, the standard for RF tooling. The simulations were performed using the \emph{multi-modal eigenfrequency} calculation, followed by a \emph{modal frequency sweep} that constructs the filter response based on the determined eigenfrequencies. This method is much faster then a exhaustive frequency sweep. This enables a larger number of iterations at a faster rate. The simulations were set to search for a number of eigenfrequencies greater than the number of poles of the designed filter. Each filter that was manufactured was first verified using COMSOL. The results of these simulations are presented below in Figures \ref{figure:simCOMSOL}a and \ref{figure:simCOMSOL}b. Notice that the simulations yield the expected passband for the proposed geometries, with slight differences. For the interdigital stage the simulated passband was slightly lower than designed, and for the parallel stage the breadth of the passband was slightly shorther. In particular, several variations of the parallel design presented difficulties around broadening the passband. The artefacts at higher frequencies were ignored as there should not be any frequency content in those areas during normal operation.

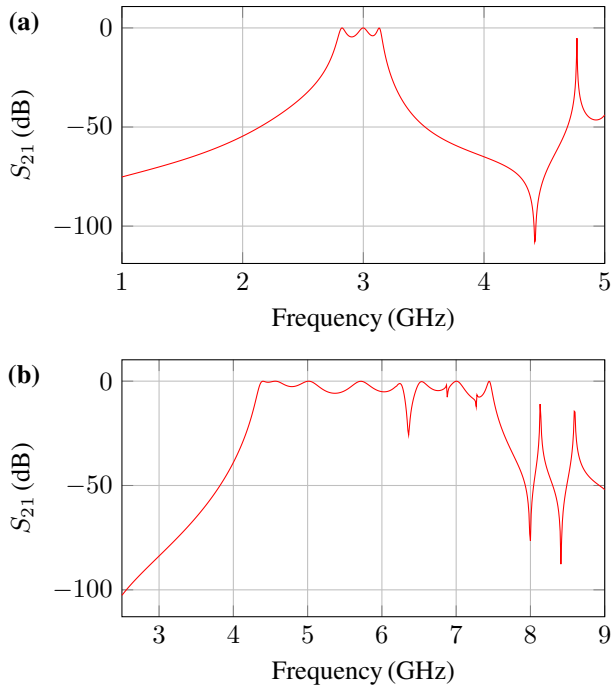
\begin{figure}[htb]
\centering
\begin{tikzpicture}
	\begin{axis}[
		height=5cm,
		width=8cm,
		xlabel={Frequency\,(GHz)},
		ylabel={$S_{21}$\,(dB)},
		xmin=1,xmax=5,
		grid
		]
		\addplot[red] table[x index = {0}, y index = {1}, col sep=comma] {figSimID12.csv};
	\end{axis}
	\begin{axis}[
		yshift=-4.7cm,
		height=5cm,
		width=8cm,
		xlabel={Frequency\,(GHz)},
		ylabel={$S_{21}$\,(dB)},
		xmin=2.5,xmax=9,
		grid
		]
		\addplot[red] table[x index = {0}, y index = {1}, col sep=comma] {figSimID09.csv};
	\end{axis}
\node at (-1.3,3.2) {\textbf{(a)}};
\node at (-1.3,-1.5) {\textbf{(b)}};
\end{tikzpicture}
\captionsetup{width=.8\linewidth}
\caption[Filter Simulation Results]{\textbf{Filter COMSOL simulation results}.  Simulations were performed seeking 8 eigenfrequencies. \textbf{(a)} Interdigital filter simulation for parameters in Table~\ref{table:ID12_params}. Width of the passband lies between 2.75\,GHz and 3.25\,GHz as required for the first stage. The stopband has high frequency resonances after 4.5\,GHz. \textbf{(b)} Parallel coupled filter for parameters in Table~\ref{table:ID09_params}. Passband shows undesirable resonant modes between 6 and 6.5\,GHz. Overall width of the passband is noticeably large making it ideal for the second stage. Several iterations of parallel filters proved difficult to push the width of the passband beyond that presented.} 
\label{figure:simCOMSOL}
\end{figure}

\subsection{Manufactured Filter Evaluation}

\begin{figure}[htb]
\centering
\input{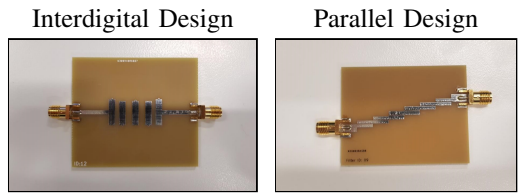}
\captionsetup{width=.8\linewidth}
\caption[Manufactured microstrip filters used for double upconversion]{\textbf{Manufactured microstrip filters used for double upconversion}. Photos show top layer of filters on standard FR4 dielectric. The bottom layers are full ground planes.}
\label{figure:manufactured_filts}
\end{figure}

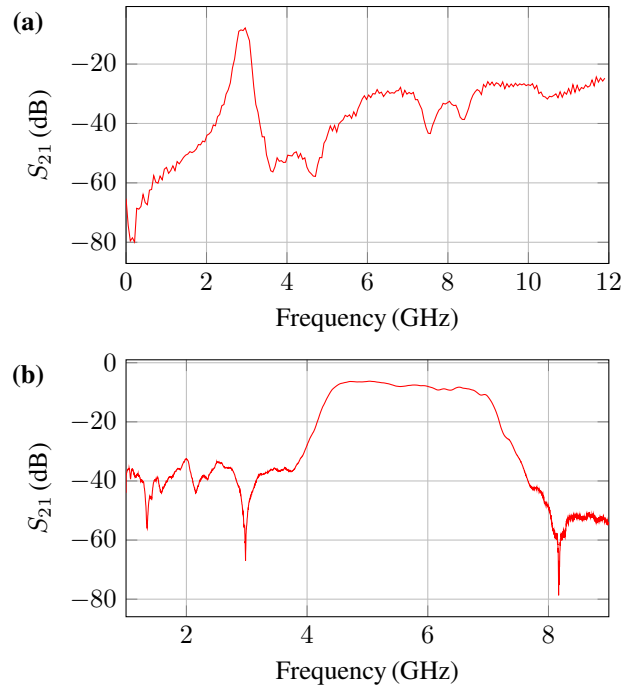
\begin{figure}[htb]
\centering
\begin{tikzpicture}
	\begin{axis}[
		height=5cm,
		width=8cm,
		xlabel={Frequency\,(GHz)},
		ylabel={$S_{21}$\,(dB)},
		xmin=0,xmax=12,
		grid
		]
		\addplot[red] table[x index = {0}, y index = {1}, col sep=comma] {figNAid12.csv};
	\end{axis}
	\begin{axis}[
		yshift=-4.7cm,
		height=5cm,
		width=8cm,
		xlabel={Frequency\,(GHz)},
		ylabel={$S_{21}$\,(dB)},
		xmin=1,xmax=9,
		grid
		]
		\addplot[red] table[x index = {0}, y index = {1}, col sep=comma] {figNAid09gndunsol.csv};
	\end{axis}
	\node at (-1.3,3.2) {\textbf{(a)}};
	\node at (-1.3,-1.5) {\textbf{(b)}};
\end{tikzpicture}
\captionsetup{width=.8\linewidth}
\caption[Geometry Magnitude Results]{\textbf{Transmission response of manufactured PCB filters}. The designs are that for simulations shown in Figure~\ref{figure:simCOMSOL}. \textbf{(a)} Transmission of first stage interdigital filter designed using a fifth order Butterworth polynomial. Passband lies roughly between 2.8\,GHz and 3\,GHz with attenuation on average of 8\,dB. \textbf{(b)} Transmission of second stage parallel filter designed using a fifth order Butterworth polynomial. Passband lies roughly between 4.5\,GHz and 7\,GHz with attenuation on average of 5\,dB.}
\label{figure:measNAfilters}
\end{figure}

The fabricated filters used are presented in figure \ref{figure:manufactured_filts}. Once fabricated, a filter's response (scattering parameters) were measured using a vector network analyser. The magnitude transmission responses ($S_{21}$) of the filters are presented in Figures \ref{figure:measNAfilters}a and \ref{figure:measNAfilters}b. 

The response of the interdigital filter presented in Figure~\ref{figure:measNAfilters}a shows a narrow passband between 2.8\,GHz and 3\,GHz with fast roll-off either side. The attenuation in the passband averages around 8dB. Given this response, LO1 will feed the first mixer to place the IF signal at 2.9\,GHz. This will result in a spacing of 2.9\,GHz of the upconverted IF signal and the frequency of LO2 after the second mixer stage. 

The parallel design shows a desirable bandpass response with a passband 2.5\,GHz in width. Compared to the stage 1 filters, the attenuation in the passband is smaller, having a minimum attenuation of approximately 4\,dB.

\subsection{Double Up Conversion Circuit}

\begin{figure}[htb]
\centering
\includegraphics[scale = 0.26]{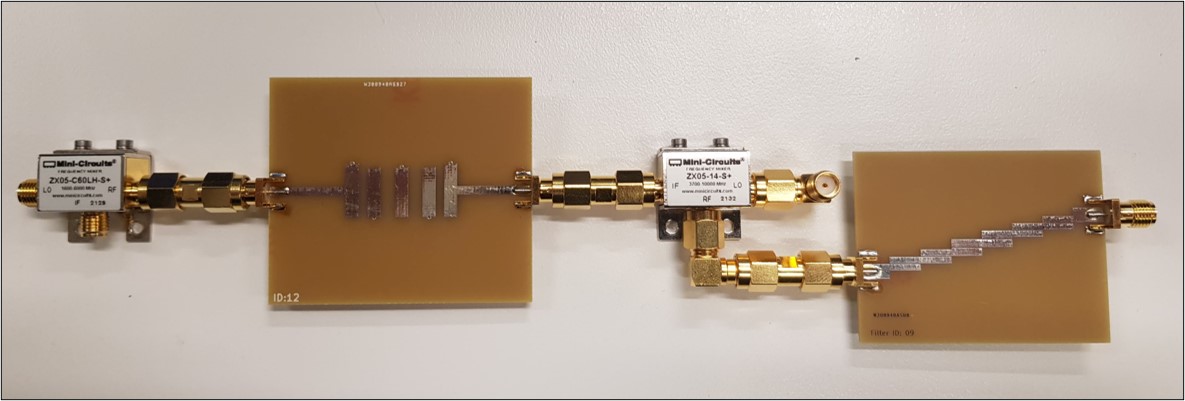}
\captionsetup{width=.8\linewidth}
\caption[Double Up Conversion Construction]{\textbf{Physical realisation of single qubit circuit} using two mini circuits mixers and interdigital and parallel geometry filters for first and second stage respectively.}
\label{figure:single_construction}
\end{figure}

Given two viable filter designs that had been manufactured, a double upconversion circuit was built using 2 low-cost \emph{Mini Circuits} mixers \emph{ZX05-C60LH-S+} and \emph{ZX05-14}. The result is the device presented in Figure~\ref{figure:single_construction} which is a physical realisation of the design presented in Figure~\ref{figure:double_up_topology}.

\section{Results}

\subsection{Qubit Control with Double Conversion} \label{double_up_conversion}
In order to control the qubit utilising the double up conversion technique, the circuit presented in figure \ref{figure:single_construction} was wired up with an AWG source to the first mixer input, and two separate LO sources were wired to the first and second mixer. The output of the second stage filter was then wired to the drive input port of the qubit at the interface of the fridge housing the superconducting circuit. The first LO source was set to a constant frequency (2.5\,GHz in this case), placing the lower sideband copy of the drive signal within the passband of the first stage filter. A single AWG source was then programmed to output a constant sinusoidal drive at a frequency of 500\,MHz what would ensure that the upmixing operation would place it in the passband of the first stage filter at 2.9\,GHz.
Once both local oscillators powering the single qubit device were set to place the AWG IF signal at the qubit frequency, a Rabi experiment was performed as described in ~\cite{Krantz2019,bais,PhysRevApplied.16.054039}. The results of the experiment are presented below in Figure~\ref{figure:rabi_with_module}. The measurement is a standard experiment of a transmon qubit in a dry dilution fridge.

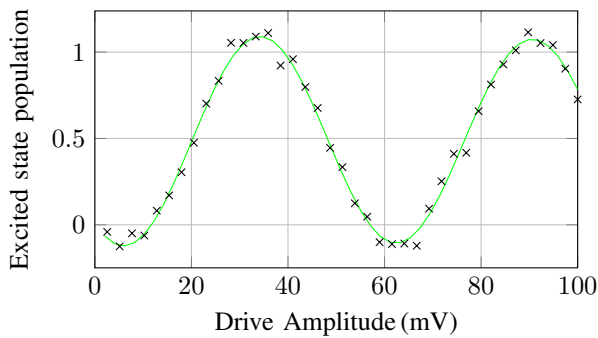
\begin{figure}[htb]
\centering
\begin{tikzpicture}
	\begin{axis}[
		height=5cm,
		width=8cm,
		xlabel={Drive Amplitude\,(mV)},
		ylabel={Excited state population},
		xmin=0,xmax=100,
		grid
		]
		\addplot [only marks, mark=x] table[x index = {0}, y index = {1}, col sep=comma, skip first n=1] {figExpRabiRaw.csv};
		\addplot [green] table[x index = {0}, y index = {1}, col sep=comma, skip first n=1] {figExpRabiFit.csv};
	\end{axis}
\end{tikzpicture}
\captionsetup{width=.8\linewidth}
\caption[Rabi Experiment Results with Double up conversion Circuit]{\textbf{Results of a Rabi experiment using double upconversion circuit}. The drive pulse was held at a constant of 50\,ns while the drive amplitude was varied. Results show a successful characterization of oscillation between ground and excited states as the drive pulse rotates the qubit state about the x-axis.}
\label{figure:rabi_with_module}
\end{figure}

The results presented in Figure \ref{figure:rabi_with_module} display that the double upconversion technique implemented with the single qubit module was able to successfully invoke x-axis rotations of the qubit state about the Bloch sphere. The fitting of the sinusoid shows that the dynamics of the qubit's state match the expected Hamiltonian of the transmon~\cite{bais}. The results of this experiment allow for drive pulses to be calibrated based on the drive amplitudes required to rotate the state vector about the x-axis by a certain amount. Further experimentations validating the circuit are presented in appendix~\ref{app:control_with_double}

\subsection{Analysis of Limitations}\label{sect:limitations}

It is important that the utilisation of the device takes into account the non-ideal components in use. The mixers are not capable of up-mixing any arbitrary inputs. In particular, the chosen second stage mixer is only specified to function for LO inputs between 4\,GHz and 10\,GHz. As a result the second mix up stage must always place the LO above the passband of the second stage filter. However, as the spacing between the second LO and drive sidebands is 2.9\,GHz the drive signal can still be placed anywhere in the 2.5\,GHz range of the second filter's passband while still attenuating the second LO. Essentially, as the first stage filter allows for upconversion to a high enough frequency, capability is provided to sweep the lower sideband drive signal of the second upconversion across the full range of the second filter's passband of 4.5 to 7\,GHz.  

\subsubsection{Spectra of Double Upconversion}\label{sect:spectrums of qubit module}

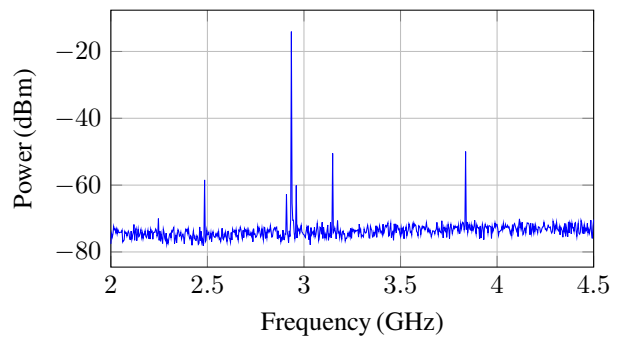
\begin{figure}[htb]
\centering
\begin{tikzpicture}
	\begin{axis}[
		height=5cm,
		width=8cm,
		xlabel={Frequency\,(GHz)},
		ylabel={Power\,(dBm)},
		xmin=2,xmax=4.5,
		grid
		]
		\addplot[blue] table[x index = {0}, y index = {1}, col sep=comma] {figMeasSpecFirst.csv};
	\end{axis}
\end{tikzpicture}
\captionsetup{width=.8\linewidth}
\caption[First up conversion spectrum]{\textbf{Measured Spectrum after the first stage up conversion of a double up conversion circuit}. Desired signal is noticeable at 2.9GHZ. Some spurious peaks are noticeable around the desired sideband. RBW for measurements set to 1\,MHz. The power of all sources was set to 10\,dBm. The input source was set to 450\,MHz with the first and second LO being set to 3.35\,GHz and 7.926\,GHz respectively.}
\label{figure:first up conversion spec}
\end{figure}

\begin{figure}[htb]
\centering
\begin{tikzpicture}
	\begin{axis}[
		height=5cm,
		width=8cm,
		xlabel={Frequency\,(GHz)},
		ylabel={Power\,(dBm)},
		xmin=4.5,xmax=6,
		grid
		]
		\addplot[blue] table[x index = {0}, y index = {1}, col sep=comma] {figMeasSpecSecond.csv};
	\end{axis}
\end{tikzpicture}
\captionsetup{width=.8\linewidth}
\caption[Second up conversion spectrum]{\textbf{Measured Spectrum at the output of a double up conversion circuit}. The attenuated LO and upper sideband from the first stage up conversion are noticeable to the right of the desired signal. RBW for measurements was set to 1\,MHz. The power of all sources was set to 10~dBm. The input source was set to 450\,MHz with the first and second LO being set to 3.35\,GHz and 7.926\,GHz respectively.}
\label{figure:second up conversion spec}
\end{figure}
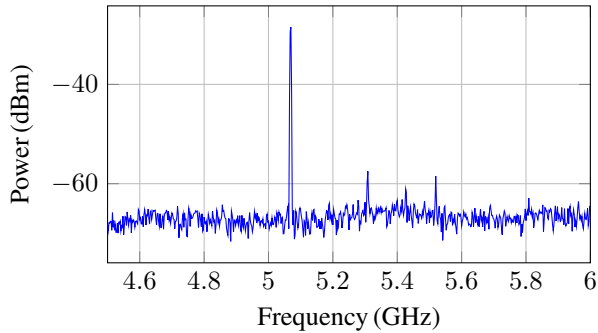

To develop a thorough evaluation of the circuits performance, the power spectra after each upconversion stage were analysed. Figures \ref{figure:first up conversion spec} and \ref{figure:second up conversion spec} show the spectra after the first and second stages of the double upconversion respectively. The setup used to capture these spectra consisted of 3 RF sources provided by \emph{WindFreak SynthHD Pro (v2)}; one for the input IF and two for the LO stages. The power of all sources was set to 10dBm. The input source was set to 450\,MHz with the first and second LO being set to 3.35\,GHz and 7.926\,GHz respectively. 

Figure~\ref{figure:first up conversion spec} shows the desired upconverted sideband signal at 2.9\,GHz. Analysis of this spectrum reveals a limiting factor of the circuit introduced by the first stage filter. The second stage mixing process will also up convert the attenuated first stage LO and unwanted sideband alongside the desired IF signal. Given that all of these signals are relatively close in frequency, in a majority of cases they will all lie within the passband of the second stage filter. As such, the attenuation of the first stage LO and undesirable sideband is decided predominantly by the response of the first stage filter. In this scenario there is a separation between the desired sideband and undesired sideband of approximately 42\,dBm. Also of note in Figure~\ref{figure:first up conversion spec} is the unwanted spectral peaks on the lower side of the desired sideband. These spectral peaks were difficult to recreate consistently and quite spurious in their nature, a further investigation of them is available in Section \ref{sect:antenna}.

Figure \ref{figure:second up conversion spec} shows the output of the double upconversion circuit. As discussed above the attenuation of the first LO and first stage undesired sideband is a result of the first stage filtering process only, and as such both spectral peaks are still present in this signal. In this case the minimum separation between desired and undesirable signal is approximately 30~dB. The reasoning for this difference in separations between first and second up conversion is due to the differing roll-off of the second stage filter. In comparison to Figure~\ref{figure:first up conversion spec}, there are no major significant spurious spectral peaks. 

The spectra presented in Figures \ref{figure:first up conversion spec} and \ref{figure:second up conversion spec} show strong separation between desired frequencies and unwanted spectral peaks. In comparison, IQ modulation within the lab can achieve attenuation's ranging between 30 and 40\,dBm.

\begin{figure}[htb]
\centering
\begin{tikzpicture}
	\begin{axis}[
		height=5cm,
		width=8cm,
		xlabel={Frequency\,(GHz)},
		ylabel={Power\,(dBm)},
		xmin=4.5,xmax=8.5,
		grid
		]
		\addplot[blue] table[x index = {0}, y index = {1}, col sep=comma] {figMeasSpecDBc.csv};
	\end{axis}
\end{tikzpicture}
\captionsetup{width=.8\linewidth}
\caption[Spectrum Measurement for dBc]{\textbf{Spectrum used to compute dBc}. Example spectrum used to compute dBc of LO2 power relative to up converted drive signal. The second LO signal is noticeable several GHz above the desired signal. dBc is computed as the difference in dBm between LO2 and desired signal. RBW for measurements was set to 1\,MHz}
\label{figure:dBc spectrum}
\end{figure}
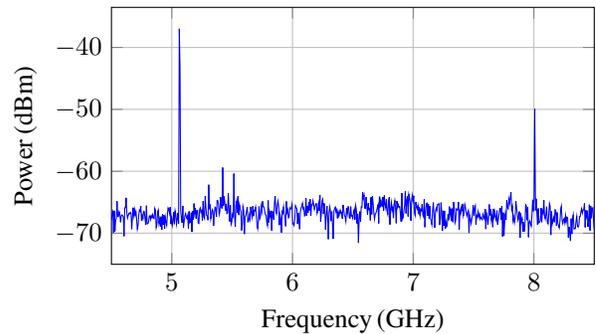

\subsubsection{Benchmarking performance} \label{benchmark}

Although it was demonstrated that this circuit can adequately control a qubit, detailed analysis of the spurious peaks were done across the range of potential qubit frequencies. Additionally, measurements were performed on a set of 4 identical circuits to gauge reproducibility. To evaluate the performance, we used decibels relative to the desired signal (dBc). This metric compares the power of the drive signal at the qubit frequency to the power of second local oscillator, providing a measure of isolation of the drive signal. An example spectrum that was captured in the measurement process is presented in Figure~\ref{figure:dBc spectrum}. This figure highlights the limitations of the first filter attenuation as the LO1 and undesired first stage sideband are still present in the signal (refer to Section~\ref{sect:spectrums of qubit module} for further details). More importantly, the second LO in this figure is showing considerable power despite expected attenuation of the second filter. The dBc measure in this case will reference the second LO signal as it is the second most powerful signal. 

\begin{figure}[ht]
\centering
\begin{tikzpicture}
	\begin{axis}[
		height=5cm,
		width=8cm,
		xlabel={Frequency\,(GHz)},
		ylabel={Power\,(dBc)},
		xmin=4.5,xmax=6.867,
		grid
		]
		\addplot table[x index = {0}, y index = {1}, col sep=comma, skip first n=1] {figDBCmultiple.csv};
		\addplot table[x index = {0}, y index = {2}, col sep=comma, skip first n=1] {figDBCmultiple.csv};
		\addplot table[x index = {0}, y index = {3}, col sep=comma, skip first n=1] {figDBCmultiple.csv};
		\addplot table[x index = {0}, y index = {4}, col sep=comma, skip first n=1] {figDBCmultiple.csv};
	\end{axis}
\end{tikzpicture}
\captionsetup{width=.8\linewidth}
\caption[dBc Drive to LO2 Benchmark of Double up Conversion Circuits]{\textbf{dBc measurements of multiple double up conversion circuits}. dBc measurements taken between up converted drive signals and second stage LO across a range of frequencies within the passband of the second stage filter.}
\label{figure:dBc}
\end{figure}
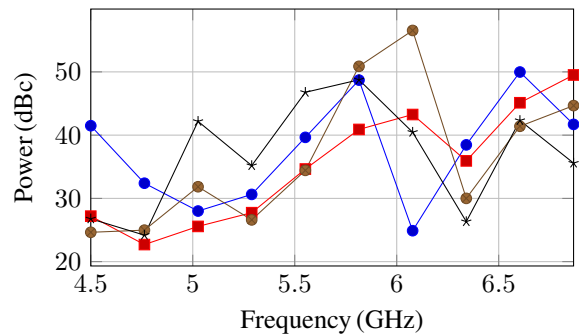

All measurements were taken using an AWG drive signal set to 450\,MHz, a frequency that is available to all AWG's typically present in a superconducting quantum devices lab. The first LO was set to 3.35\,GHz to place the drive signal at 2.9\,GHz in the passband of the first stage filter. \emph{WindFreak} RF sources were used for both LO inputs as well as the representative drive IF signal. All sources were set to 10\,dBm power. The dBc was then recorded across a range of frequencies that sit in the passband of the second stage filter by shifting the frequency of the second LO. The results are presented in Figure~\ref{figure:dBc}.

An IQ modulation scheme can often achieve between 30 to 40\,dBc between carrier and drive signal. Reviewing Figure~\ref{figure:dBc} shows that a majority of the measurements of the double upconversion circuits sits within this range, with some certain frequencies showing worse performance and others showing an out performance of IQ modulation attenuation.   

It is important to note the variation in spacing of LO frequencies between IQ modulation and double upconversion. In the case of IQ modulation, the dBc of LO frequencies to drive signal is of two signals spaced only by the frequency of drive signal, often restricted to the order of 100\,MHz. In comparison, double upconversion has the LO spaced much further from the drive, in the order of GHz; in this case 2.9\,GHz. This distancing is much more desirable as it is far less likely to cause unwanted dynamics as the LO signal is very off-resonant to the qubit. 

Regardless of this larger separation provided by the double up conversion, the limitations determined by the first stage filter presented in Section~\ref{sect:spectrums of qubit module} remain a determining factor for the quality of the drive signal. It is important to note that in scenarios above where dBc relative to the second LO outperforms IQ modulation, the dBc relative to the first LO remains constrained by the first filter; on par with performance of IQ modulation.

Finally, there are small peaks seen after the desired frequency tone in Figure~\ref{figure:dBc spectrum} due to harmonics of LO1 and the RF tone generated from the first stage mixer. These can be eliminated by adding a lowpass filter after 4-5\,GHz after the first stage bandpass filter to enhance the stopband shown in Figure~\ref{figure:measNAfilters}a. Note that this can be implemented in-situ on the filter PCB by adding a notch filter after the interdigital stage like a butterfly stub~\cite{Hong2004-pn}.

\subsubsection{Filter's acting as antennas}\label{sect:antenna}

Despite adequate results shown in Figure~\ref{figure:dBc}, it was observed that the second stage filter was not attenuating the LO2 signal as expected based on the filter's response presented in Figure~\ref{figure:measNAfilters}b. As the filters are made of microstrip copper open to the air, it is possible that the filters are acting as antennas to other microwave fields generated within the circuit. To prove that the filter was indeed acting as an antenna, it was connected to the spectrum analyser roughly 500\,cm from the rest of the circuit. Measurements were taken and the results are presented below in Figure~\ref{figure:antenna}. Note that these measurements are simply qualitative and will vary with the given environment; nonetheless, they are presented to convey evidence of the antenna-like properties.

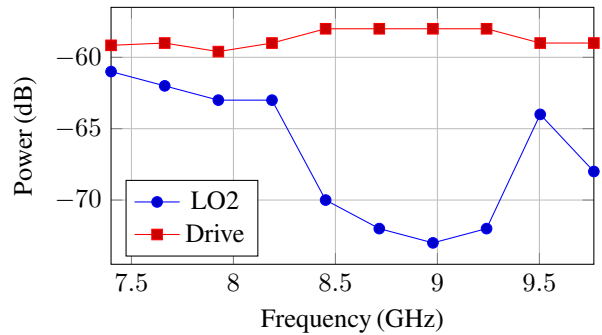
\begin{figure}[ht]
\centering
\begin{tikzpicture}
	\begin{axis}[
		height=5cm,
		width=8cm,
		xlabel={Frequency\,(GHz)},
		ylabel={Power\,(dB)},
		xmin=7.4,xmax=9.767,
		legend pos=south west,
		grid
		]
		\addplot table[x index = {0}, y index = {2}, col sep=comma, skip first n=1] {figDBCantenna.csv};
		\addlegendentry{LO2}
		\addplot table[x index = {0}, y index = {3}, col sep=comma, skip first n=1] {figDBCantenna.csv};
		\addlegendentry{Drive}
	\end{axis}
\end{tikzpicture}
\captionsetup{width=.8\linewidth}
\caption[Disconnected Filter Measurements]{\textbf{Measurements utilising second stage filter as an antenna}. Measurements taken with second stage filter removed from double up conversion circuit and separated by approximately 500cm. Input power was set to 10dBm. Results show that the filter is acting as an antenna for both the first stage up converted signal and the second stage LO.}
\label{figure:antenna}
\end{figure}

Figure~\ref{figure:antenna} shows that not only the LO2 signal but the drive signal after first stage mixer are being picked up by the second stage filter a considerable distance away. Other second stage filter designs showed nearly identical results. Nonetheless, the successful control of the qubit presented in Section~\ref{double_up_conversion} was completed using the unshielded variation presented in this paper. However, RF shielding is recommended given that qubits controlled at other frequencies could be more vulnerable to the unwanted spectral peaks introduced by the filter acting as an antenna. The result of such shielding should increase dBc across all frequencies. Furthermore, the variation in dBc  across devices presented in Figure~\ref{figure:dBc} would also be reduced as erroneous over the air frequencies will no longer resonate with the filter.

\begin{figure}[htb]
\centering
\begin{tikzpicture}
 	\definecolor{leGreen}{rgb}{0.03529, 0.73333, 0.623529}
	\definecolor{leOrange}{rgb}{0.996078, 0.7098, 0.392549}
	\begin{axis}[
		height=5cm,
		width=8cm,
		xlabel={Frequency\,(GHz)},
		ylabel={Power\,(dBm)},
		xmin=1,xmax=9,
		legend pos=north west,
		grid
		]
		\addplot[blue] table[x index = {0}, y index = {1}, col sep=comma] {figSpecFinalA.csv};
		\addlegendentry{5\,GHz}
		\addplot[red,draw opacity=0.83] table[x index = {0}, y index = {1}, col sep=comma] {figSpecFinalB.csv};
		\addlegendentry{6\,GHz}
	\end{axis}
\end{tikzpicture}
\captionsetup{width=.8\linewidth}
\caption[First up conversion spectrum]{\textbf{Pure tones generated when using RF shielding}. Two spectra (taken with 1\,kHz RBW) in a typical qubit frequency range are shown: 5\,GHz and 6\,GHz. For both cases, the IF drive frequency was 460\,MHz at 0\,dBm power, the LO1 frequency was 2.95\,GHz at 10\,dBm power and the LO2 power was 10\,dBm. There are no spurious peaks seen over a 70~dB range.}
\label{figure:finalSpec}
\end{figure}
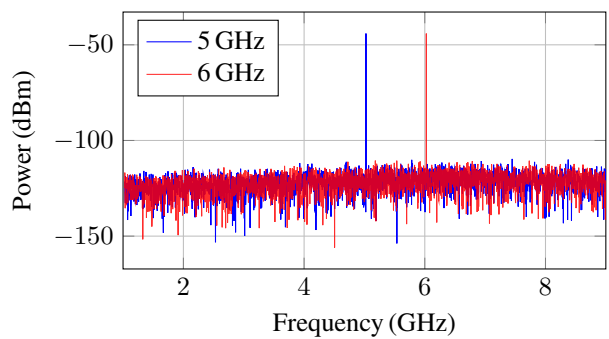

Although the scheme in this presented form is sufficient for the control of superconducting qubits, we push the limits of this setup in Figure~\ref{figure:finalSpec}. Here we placed each filter inside separate aluminium cans sealed with aluminium tape to provide RF shielding. In addition, the first stage filter was augmented with another identical interdigital PCB filter along with two \emph{Mini Circuits} \emph{VLF-5000+} filters to enhance the stopband beyond 5\,GHz. Two tones (5 and 6\,GHz) were generated in the usual qubit range. These tones were separately generated using the same IF and LO1 inputs while only changing the frequency of LO2. The result is at least 70\,dB suppression of all spurious peaks over the relevant control range of the transmon qubit from 1 to 9\,GHz.

\section{Conclusion}

The utilisation of a double upconversion scheme enables a low-cost alternative for synthesising RF signals at frequencies required for control of superconducting qubits. Manufacturing of microstrip filters on standarad PCB dielectrics proved feasible for filtering out unwanted artefacts of each upconversion process within the scheme. Despite a demonstration of successful qubit control with equivalent performance to conventional IQ modulation, there are significant improvements that can be made to get greater control fidelities: specifically optimising the filter roll-offs and shielding the circuit components from RF crosstalk. Nonetheless, the scheme is more scalable than IQ modulation in terms of cost as each qubit now only requires a single expensive AWG output (unlike two in IQ modulation) while compromising with two (cheaper) RF sources for the LO inputs instead of one. In addition, the scheme does not require a specialised IQ mixer or any in-situ calibration when operating a qubit. Finally, we showed that when using proper shielding, the qubit tones can be generated with over 70\,dB of spurious-free dynamic range over the entire operational frequency range of a transmon qubit.

\section{Acknowledgements}

The experimental design was conceived by P. Pakkiam. Simulation, optimisation and implementation were done by J. Dearlove, while measurements were jointly conducted by J. Dearlove and P. Pakkiam.

The work was supported by the Australian Research Council Centre of Excellence for Engineered Quantum Systems (EQUS, CE170100009) and by grant number FQXi-IAF19-04 from the Foundational Questions Institute Fund, a donor advised fund of Silicon Valley Community Foundation.


%

\appendices

\section{Original First Stage Filter Magnitude Response}\label{appendix:id07}

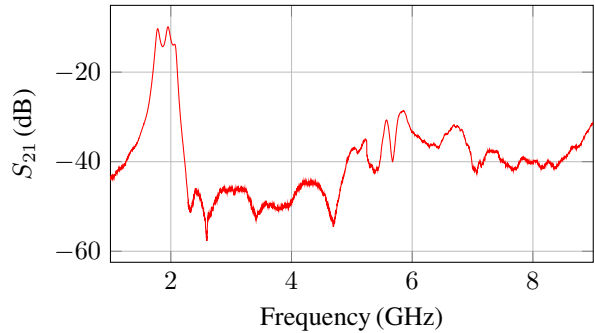
\begin{figure}[htb]
\centering
\begin{tikzpicture}
	\begin{axis}[
		height=5cm,
		width=8cm,
		xlabel={Frequency\,(GHz)},
		ylabel={$S_{21}$\,(dB)},
		xmin=1,xmax=9,
		grid
		]
		\addplot[red] table[x index = {0}, y index = {1}, col sep=comma] {figNAid07gndunsol.csv};
	\end{axis}
\end{tikzpicture}
\captionsetup{width=.8\linewidth}
\caption[Original First Stage Filter Magnitude Results]{\textbf{Amplitude ($S_{21}$) response of initial first stage filter} designed using a 5th order Chebyshev polynomial. Passband lies roughly between 1.8\,GHz and 2.2\,GHz.}
\label{figure:id07mag}
\end{figure}

Figure~\ref{figure:id07mag} is the response of the filter used to control the qubit as per the results presented in figure ~\ref{figure:rabi_with_module}. All other results presented were created using the filter with response as presented in figure~\ref{figure:measNAfilters}a.

\section{Qubit Control with Double Upconversion}\label{app:control_with_double}

This section will break down the utilisation of double up conversion to successfully characterize and control a transmon qubit. 
This section will also draw comparisons to how these operations would be performed with a setup that requires IQ modulation for synthesising qubit drive signals. The results presented below are pivotal in verifying that the work presented can be used as a valid replacement for IQ mixing for creating qubit drive signals. The following experiments were all performed on transmon qubit. The qubit was flux tuned to the "sweet spot" for all operations \cite{roth_ma_chew_2021}. 

\subsubsection{Qubit Frequency Characterization} \label{sect:qubit freq chara}

Before a qubit can be operated, its frequency must be characterised. In order to identify the qubit frequency, a constant amplitude drive is applied and the frequency is swept over a range of potential qubit frequencies. 

In order to complete this frequency sweep utilising double upconversion, the circuit was wired up with an AWG source to the first mixer input, and two separate LO sources were wired to the first and second mixer. The output of the second stage filter was then wired to the drive input port of the qubit at the interface of the fridge. The first LO source was set to a constant frequency (2.5\,GHz in this case), placing the lower side band copy of the drive signal within the passband of the first stage filter. A single AWG source was then programmed to output a constant sinusoidal drive at a frequency that would ensure that the up-mixing operation would place it in the passband of the first stage filter. Once this was set, the second LO was swept across the range of expected qubit frequencies. Results are presented below in Figure \ref{figure:freq_chara_with_module}. 

\begin{figure}[htb]
\centering
\begin{tikzpicture}
	\begin{axis}[
		height=5cm,
		width=8cm,
		xlabel={Drive Frequency\,(GHz)},
		ylabel={IQ Amplitude (a.u.)},
		xmin=5.54999999,xmax=5.649,
		grid
		]
		\addplot [only marks, mark=x] table[x index = {0}, y index = {1}, col sep=comma, skip first n=1] {figExpQubitSpecRaw.csv};
		\addplot [green] table[x index = {0}, y index = {1}, col sep=comma, skip first n=1] {figExpQubitSpecFit.csv};
	\end{axis}
\end{tikzpicture}
\captionsetup{width=.8\linewidth}
\caption[Qubit Frequency Characterization with Double Upconversion]{\textbf{Qubit frequency characterisation completed using double upconversion}. Excitation of the qubit was identified and calibrated through fitting of a Lorentzian at 5.61\,GHz~\cite{Krantz2019}}.
\label{figure:freq_chara_with_module}
\end{figure}
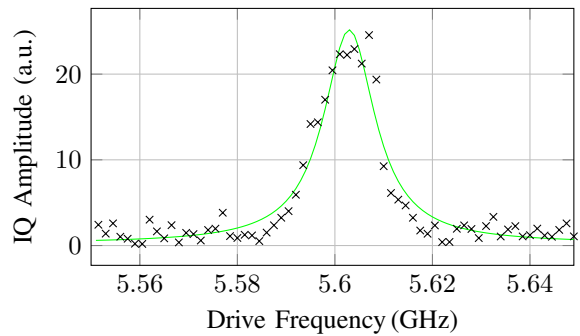

In comparison, IQ mixing could not be used to identify the qubits frequency with ease, as the mixer would need to be re-calibrated for each frequency that is scanned. As such, normal operations in the lab would require rewiring of the external electronics to perform a sweep of the potential qubit frequencies using a generic RF source which commonly are used as local oscillators. Only once this sweep has been performed and the qubit frequency identified could the control circuit be rewired to enable IQ mixing for drive signal synthesis at the qubit frequency. This need to perform physical rewiring and calibration of hardware has been removed by the integration of the double upconversion circuit.

Once the qubit frequency has been identified, the second LO is set such that any signals generated at the AWG source will be output at the qubit frequency. After the second LO is set, controlling the qubit is all handled through programming of the AWG. Following this a Rabi experiment was performed as outlined in section \ref{double_up_conversion}

\subsubsection{Ramsey Oscillations}

Following successful Rabi oscillations, a Ramsey experiment was also performed to determine the T2* (decoherence time) of the qubit \cite{coherence_decay,youssef_2020}. The utilisation of a Ramsey experiment is integral in bench marking the double up conversion method, as it will identify any potential decoherence that the method could introduce. The results of the Ramsey experiment are presented below in Figure \ref{figure:ramsey_with_module}. 

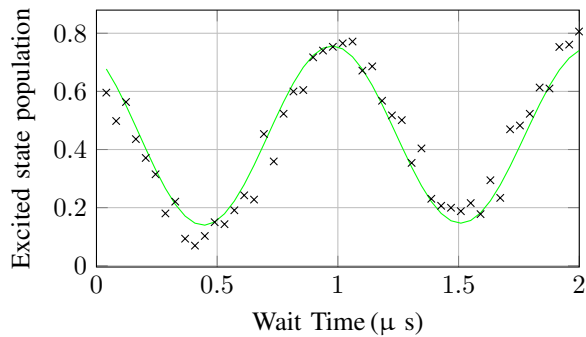
\begin{figure}[htb]
\centering
\begin{tikzpicture}
	\begin{axis}[
		height=5cm,
		width=8cm,
		xlabel={Wait Time\,($\upmu$ s)},
		ylabel={Excited state population},
		xmin=0,xmax=2,
		grid
		]
		\addplot [only marks, mark=x] table[x index = {0}, y index = {1}, col sep=comma, skip first n=1] {figExpRamseyRaw.csv};
		\addplot [green] table[x index = {0}, y index = {1}, col sep=comma, skip first n=1] {figExpRamseyFit.csv};
	\end{axis}
\end{tikzpicture}
\captionsetup{width=.8\linewidth}
\caption[Ramsey Experiment Results with Double Upconversion]{\textbf{Results of Ramsey experiment using double upconversion}. The qubit was driven with a $\pi/2$ pulse to an equal superposition on the equator of the Bloch Sphere using pulses calibrated from Rabi experiments. From here a set wait time allowed for the qubit state to evolve about the equator before another $\pi/2$ pulse was applied~\cite{Krantz2019}. Results show no noticeable phase decoherence across tested wait times.}
\label{figure:ramsey_with_module}
\end{figure}

The results presented in Figure \ref{figure:ramsey_with_module} display that no noticeable decoherence was introduced through the utilisation of a double upconversion circuit. The wait times tested were based on previous Ramsey experiments on the same qubit using IQ modulation. Following this characterization, the bench marking of the double upconversion technique has been verified. The attenuation provided by each filter stage is enough to isolate the drive signal, allowing for coherent control of the qubit that matches the performance of IQ modulation.  
\pagebreak

\ifCLASSOPTIONcaptionsoff
  \newpage
\fi



%
\bibliographystyle{ieeetr}
\bibliography{main.bib}

%






\end{document}